\documentclass[11pt]{article}

\usepackage{authblk}

\author{Gokcen Deniz Ozen \thanks{dozen@metu.edu.tr}}
\author{Sahin Kurekci \thanks{kurekci@metu.edu.tr}}
\author{Bayram Tekin \thanks{btekin@metu.edu.tr}}
\affil{Department of Physics, Middle East Technical University, 06800, Ankara, Turkey}
\date{}                     
\usepackage[utf8]{inputenc}
\usepackage[english]{babel}
\usepackage[margin=1in, paperwidth=8.5in, paperheight=11in]{geometry}
\usepackage{amsmath}
\usepackage{amssymb}
\usepackage{url}
\usepackage{relsize}
\usepackage[autostyle]{csquotes}
   \MakeOuterQuote{"}
\usepackage{soul}
\usepackage{dashrule}

\usepackage{amsthm,bm}

\usepackage{pdfpages}

\usepackage{dsfont} 

\newtheorem*{theorem*}{Theorem}
\newtheorem*{definition*}{Definition}
\newtheorem*{corollary*}{Corollary}

\newtheorem*{ex*}{Example}

\newtheorem*{remark*}{Remark}

\usepackage{tikz}

\usepackage{mathtools}

\usepackage{array,booktabs}
\newcolumntype{C}{>{$\displaystyle}c<{$}}

\usepackage{parskip} 

\usepackage{color} 

\begin{document}

\title{\textbf{Entropy in Born-Infeld Gravity}}
\date{}
\maketitle

\noindent\makebox[\linewidth]{\rule{\textwidth}{1pt}} 

\begin{abstract}
There is a class of higher derivative gravity theories that are in some sense natural  extensions of cosmological Einstein's gravity with a unique maximally symmetric classical vacuum and only a massless spin\(-2\) excitation about the vacuum and no other perturbative modes. These theories are of the Born-Infeld determinantal form. We show that the macroscopic dynamical entropy as defined by Wald for bifurcate Killing horizons in these theories are equivalent to the geometric Bekenstein-Hawking entropy  (or more properly Gibbons-Hawking entropy for the case of de Sitter spacetime) but given with an effective gravitational constant which encodes all the information about the background spacetime and the underlying theory. We also show that the higher curvature terms increase the entropy. We carry out the computations in generic \(n-\)dimensions including the particularly interesting limits of three, four and infinite number of dimensions. We also give a preliminary discussion about the black hole entropy in generic dimensions for the BI theories.
\end{abstract}

\noindent\makebox[\linewidth]{\rule{\textwidth}{1pt}} 

\tableofcontents

\noindent\makebox[\linewidth]{\rule{\textwidth}{1pt}} 

\newpage

\section{Introduction}

It is perhaps not possible to write a classical theory of gravity that can supersede Einstein's gravity in elegance and simplicity, which, in its most succinct form, reads in the absence of matter as a variational statement
\begin{equation}
\label{GRaction}
	\delta_g \int_{\mathcal{M}} d^n x \sqrt{-g} R = 0.
\end{equation}
In words, pure general relativity (GR) is a statement about the topological Lorentzian manifold \(\mathcal{M}\): \textit{finding the metrics that are critical points of the total scalar curvature on the manifold with prescribed boundary and/or asymptotic conditions to model various physical cases.} This statement is valid in generic \(n-\)dimensions but trivial for \(n=1\) and \(n=2\). For all other dimensions, pure GR boils down to searching for solutions of the non-linear partial differential equation \(R_{\mu \nu} = 0\), that is finding Ricci-flat metrics. 

In \(n=3\), the Riemann and the Ricci tensors have equal number of components (six) and so Ricci-flat metrics are locally Riemann-flat and hence there is no local gravity except at the location of the sources.  But beyond \(n>3\), the innocent looking equation (\ref{GRaction}) and its equivalent partial differential equation form are highly complicated to solve and since their first appearance 100 years ago, many fascinating predictions of these equations (perhaps the most remarkable ones being the global dynamics of the Universe and the existence of black holes and gravitational waves) have been observed. What is rather fascinating about the classical equation \(R_{\mu \nu} = 0  \) is that it leads to thermodynamics type equations such as \footnote{If Maxwell field is coupled to gravity one has a more general thermodynamical equation including the charge variation \(\phi \delta Q \) on the right-hand side where \(\phi\) is the electric potential on the horizon.}
\begin{equation}
\label{thermo1}
	\delta M = T \delta S + \Omega \delta J
\end{equation}
for black hole solutions where \(T = \frac{\kappa}{2 \pi}\) is the horizon temperature and \(\kappa\) is the surface gravity, \(\Omega\) is the angular velocity of the horizon and \(J\) is the angular momentum of the black hole while \(M\) is the mass of it. \(S\) is the entropy given by the Bekenstein-Hawking formula \cite{Bekenstein,Barden}
\begin{equation}
\label{Bek-Hawk}
S = \frac{A_H}{4G},
\end{equation}
where $A_H$ is the horizon area. The temperature \(T\) and the entropy \(S\) are defined up to a multiplicative constant which can be fixed by the help of a semi-classical analysis such as the Hawking radiation \cite{Zeldovich, HawkingRad1, HawkingRad2}. But the crucial point is that, classically \(R_{\mu \nu} = 0\) equations for a black hole encode the equation (\ref{thermo1}) which is considered as the first law of black hole thermodynamics. In fact, all four laws of black hole mechanics (or thermodynamics) follow from Einstein's equations \cite{Barden}. It was also a rather remarkable suggestion and demonstration that the arguments can be turned around and thermodynamics of Rindler horizons (with some  assumptions)  can lead to the Einstein's equations as the equation of state \cite{Jacobson}. Note that Gibbons and Hawking showed that the result  (\ref{Bek-Hawk})  is valid for the de Sitter horizons \cite{Gib_Hawk} which will be relevant for this work.

One of the current problems in gravity (or quantum gravity) research is to find a possible {\it microscopic } explanation of the entropy \(S\) in (\ref{thermo1}). Let us expound on this: from pure statistical mechanical point of view, \(S\) must be given as \(S = \ln N\) where \(N\) refers to the microstates that have the same \(M, T, \Omega\) and \(J\) values in (\ref{thermo1}).  All this says that a black hole has much more internal states than meets the eye. But classical considerations suggest that black holes have no classical hair or microscopic degrees of freedom to account for the large entropy given in (\ref{Bek-Hawk}). Therefore even though Einstein's theory yields a macroscopic thermodynamical equation such as (\ref{Bek-Hawk}), it does not seem to \textit{explain} it. Of course if the classical picture were to be correct, the four laws of black hole mechanics are not true thermodynamical equations but just a mere analogy without much deep implication. However, there is mounting evidence from various considerations (such as the microscopic degree of freedom counting for extremal black holes in string theory in terms of the D-brane charges \cite{StromingerVafa, ASen} and in 2+1 dimensional gravity in terms of the asymptotic symmetries {\it etc.}) that this is not the case and black holes do have  temperature and entropy with perhaps important implications for quantum gravity.

In trying to understand the actual content of equations (\ref{thermo1}) \& (\ref{Bek-Hawk}), it is extremely important to figure out how universal they are: namely, if the underlying theory is modified from Einstein's gravity to something else, what remains of these thermodynamical equations. Besides, we also know that both at large and short distance scales (low and high energies) GR, in its simplest form, is not sufficient to explain the observable Universe. These deficiencies are well-known: for example the accelerated expansion of the Universe and the observed age (or homogeneity and isotropy) cannot be explained by pure gravity alone, hence there are various attempts to modify it. Of course, ideally one would like to have a quantum version of gravity in order to explain the beginning of the Universe or the space and time and the singularity issues of black holes. But at the moment we do not know exactly what the symmetries, principles \textit{etc.,} of such a theory are. In any case, the idea is that Einstein's theory captures a great deal about the \textit{gravitating} Universe or subsystems (such as the solar system) but it is not yet the final word on it, hence the modifications.

In trying to go beyond Einstein's theory there are many routes to follow. One such route that we have advocated recently in several works \cite{GulluSismanTekin1, GulluSismanTekin2} is the following: let us write down all the easily detectable low energy properties of Einstein's theory (\ref{GRaction}) and try to implement them in the new modified gravity theory that we are searching for. It is quite possible that at very small distances gravity (or spacetime) will be nothing like the one proposed by Einstein's theory, so any of its low energy virtues may not survive in that regime. For example superstring theory is such an example where the spacetime or the metric of the gravity is an emergent quantity. But this does not deter us from our discussion because we shall be interested in the intermediate energy scales in string theory where Einstein's gravity appears with modified curvature corrections and the pseudo-Riemannian description of spacetime is quite accurate. 

To accomplish our goal of writing a new modified theory we may list the virtues of Einstein's theory as follows: 
\begin{itemize}
	\item[(1)] uniqueness of the vacuum: source-free GR, \(R_{\mu \nu} = 0\), has a maximally symmetric solution which is the Minkowski spacetime. Classically it is the unique vacuum and other considerations, such as the positive energy theorem, suggest that this is the lowest energy state with zero energy (under the proper decay assumptions and several other physically motivated assumptions such as the cosmic censorship, dominant energy condition and the stability of the Minkowski spacetime \textit{etc.}) 
	\item[(2)] masslessness of the spin\(-2\) graviton: about its unique vacuum if one expands the metric as \(g = \eta + h\), one observes that \(h\) obeys a massless Klein-Gordon type equation, with only transverse-traceless polarization.
	\item[(3)] diffeomorphism invariance of the theory: the spacetime is a topological Lorentzian manifold without a prior geometry before gravity is switched on and there are no preferred coordinates even with gravity switched on. 
\end{itemize}

Certainly one could add further, harder-to-see properties of the theory to this list. For example, the theory has the local causality property which can be shown in various ways such as proving that there is always  a Shapiro time-delay as opposed to a time advance. It has a \(3+1\) dimensional splitting property and a well-defined initial value (Cauchy) formulation where the full equation \(R_{\mu \nu} = 0\) can be split into constraints on a spacelike Cauchy surface and evolutions off the surface in the \textit{time} direction. But these require further assumptions on the topology of \(\mathcal{M}\) and we shall not be interested in them here. Our basic working principle will be to write down a modified version of Einstein's gravity that has the above three properties.  The third item on the list is easy to satisfy but the first two are quite hard. 

One of the earliest modifications satisfying the above requirements  was famously introduced by Einstein when he added a cosmological constant to his theory:
\begin{equation}
	\label{GRactionCC}
	\delta_g \int_{\mathcal{M}} d^n x \sqrt{-g} (R-2\Lambda) = 0.
\end{equation}
What is interesting about this theory is that, in contrast to (\ref{GRaction}), this one has a dimensionful parameter \(\Lambda\) which will set the \textit{sizes} of a gravitating system.\footnote{Note that, Newton's constant is also a dimensionful parameter that arises in the coupling of matter to gravity. Since at this stage we consider pure and classical gravity for the sake of keeping the discussion simple, we shall restore the Newton's constant below.} Now the field equations are summarized as
\begin{equation}
\label{fieldeqn_cc}
	R_{\mu \nu} = \frac{2 \Lambda}{n-2} g_{\mu \nu},
\end{equation}
where the number of dimensions \(n\) also makes an explicit appearance. All the desired properties are also valid for this theory: for \(\Lambda > 0\) de Sitter (dS) and for \(\Lambda < 0 \) anti-de Sitter (AdS) is the unique vacuum, about which there is a single massless spin\(-2\) excitation. Many solutions (such as black holes) of the \(R_{\mu \nu} = 0\) equations also solve (in modified forms) the cosmological Einstein's theory. Of course currently due to the accelerated expansion of the Universe, \(\Lambda > 0 \) is preferred. For dS, since there is a cosmological horizon, similar to the black holes, one can define a temperature and an entropy for this horizon.

Other earlier modifications of Einstein's theory are of the following form 
\begin{equation}
\label{GRactionCC}
	\delta_g \int_{\mathcal{M}} d^n x \sqrt{-g}\, (R-2\Lambda_0 + \alpha R^2 + \beta R_{\mu \nu}^2 +\gamma R_{\mu \nu \alpha \beta}^2 + \dots), 
\end{equation}
which  generically do not have the first two properties in our list: they have many different vacua, having different values of \(\Lambda \), which cannot be compared with each other and there are many massive modes besides the massless spin-2 graviton. Typically some of these modes are ghosts and so the theory is hard to make sense both at the classical and the quantum levels. 

In our search for a higher order gravity theory that has the same features as Einstein's theory, we used the ideas laid out by Deser and Gibbons \cite{DeserGibbons} who were inspired by Eddingston's proposal of using generalized volumes in the action as
\begin{equation}
\delta_{g, \Gamma} \int_{\mathcal{M}} d^n x \Big ( \sqrt{ -\det \left( g_{\mu \nu}+\gamma  R_{\mu \nu}( \Gamma)\right)} - \lambda_0 \sqrt{-g} \Big),
\end{equation}
where one assumes that the metric  $g$ and the connection $\Gamma$ are independent in this variation. 
This type of theories gained some interest recently (under the rubric \textit{Eddington inspired Born-Infeld gravity}  even though strictly speaking Born-Infeld's work in electrodynamics came much later than Eddington's work in gravity). For this path of the developments, we refer the reader to \cite{Tahsintez} and a recent review \cite{Jimenez}. Instead, Deser-Gibbons's suggestion was to consider actions of the type 
 \begin{equation}
\label{GD}
\delta_{g} \int_{\mathcal{M}} d^n x  \sqrt{ -\det (  g_{\mu \nu} + \gamma R_{\mu \nu}(g)+\cdots ) }, 
\end{equation}
where the connection is the unique Levi-Civita connection, not to be varied independently. 
Of course the resulting theories are quite different: the action  (\ref{GD}) generically is not unitary nor does it have a unique vacuum. Only recently, the most general theory satisfying these requirements was found in  \cite{BINMG1,BINMG2} for \(3-\)dimensions 
and in \cite{GulluSismanTekin1, GulluSismanTekin2} for four and higher dimensions.  The \(3-\)dimensional theory is particularly elegant as it is formed by taking the determinant of the Einstein tensor plus the metric
\begin{equation}
\label{3d_action}
	I = - \frac{m^2}{4\pi G_3}\int_{\mathcal{M}} d^3 x \left[ \sqrt{-\det \left( \mathsf{g} + \frac{\sigma}{m^2} \mathsf{G}  \right)} - \left( 1 - \frac{\lambda_0}{2}    \right)\sqrt{-\mathsf{g}}   \right],
\end{equation}
with \( G_{\mu \nu} = R_{\mu \nu} - \frac{1}{2}g_{\mu \nu} R    \). Here $\sigma^2 =1$ and the theory has a  unique maximally symmetric vacuum with an effective cosmological constant given as 
\begin{equation}
\Lambda = \sigma m^2 \lambda_0  \Big ( 1- \frac{\lambda_0}{2} \Big), \quad \lambda_0 \neq 2 
\end{equation}
and a massive spin-2 excitation about this vacuum with a mass $M_g^2 = m^2 + \Lambda$.  The Wald entropy, the $c-$function and the  $c-$charge of this theory were studied in \cite{GulluSismanTekin2}, but  to set the stage for the generic Born-Infeld (BI) gravity in higher dimensions,  we reproduce the results and give some more details on the computations and also compute the second Born-Infeld type extension of new massive gravity in \(3-\)dimensions, which was not computed before.

For higher dimensions, the simple form (\ref{3d_action}) does not lead to a unitary theory, therefore one needs to add at least quadratic terms in curvature inside the determinant. After a rather tedious computation laid out in detail in \cite{GulluSismanTekin1, GulluSismanTekin2} the following theory was found:
\begin{equation}
\label{L3-2}
	I = \frac{1}{8\pi \gamma G} \int_{\mathcal{M}} d^n x \sqrt{-g} \left(  \sqrt{\det (\delta^{\rho}_{\nu}+ \gamma A^{\rho}_{\nu})} - \lambda_0 -1 \right),
\end{equation}
where the \(\mathsf{A}-\)tensor reads
\begin{align}
\label{Amunu}
A_{\mu \nu} = R_{\mu\nu}&+\beta S_{\mu\nu} +\gamma \left(a_{1}C_{\mu\rho\sigma\lambda}C_{\nu}{}^{\rho\sigma\lambda}+a_{3}R_{\mu\rho}R^{\rho}_{\nu} \right)+\dfrac{\gamma}{n}g_{\mu\nu} \left(b_{1}C_{\alpha \beta \rho \sigma}^2 +b_{2}R_{\rho \sigma}^2   \right).
\end{align}
Here, \( S_{\mu \nu} = R_{\mu \nu} - \frac{1}{n}g_{\mu \nu}R   \) is the traceless-Ricci tensor and \( C_{\mu\rho\sigma\lambda}  \) is the Weyl tensor given as
\begin{equation}
	C_{\mu\rho\sigma\lambda} = R_{\mu\rho\sigma\lambda} + \frac{2}{n-2}\left(  g_{\rho [\sigma} R_{\lambda] \mu} - g_{\mu [\sigma} R_{\lambda] \rho}   \right) + \frac{2R}{(n-1)(n-2)}  g_{\mu [\sigma} g_{\lambda] \rho}.  
\end{equation}
Observe that there are three other possible quadratic corrections \( C_{\mu \rho \nu \sigma}R^{\rho \sigma}   \), \(S_{\mu \rho}S^{\rho}_{\nu}\) and \(S_{\rho \sigma}^2\) which do not survive since they lead to massive particles and/or destroy the uniqueness of the vacuum.

The theory has five dimensionless parameters \(\beta, a_i, b_i\) and a dimensionful one which is the Born-Infeld parameter \(\gamma\) with dimensions of \(L^2\). Unitarity of the excitations of the theory and the requirement that the theory has a unique viable vacuum with only one massless spin\(-2\) graviton about this vacuum leads to a further elimination of three of these parameters. Namely, the following constraints must be satisfied
\begin{equation}
\label{constraints}
a_1 + b_1 = \frac{(n-1)^2}{4 (n-2) (n-3)}, \qquad a_3 = \frac{\beta + 1}{4}, \qquad b_2 = - \frac{\beta}{4}.
\end{equation}
Therefore after all the constraints are applied, there are only two free dimensionless parameters one of which can be chosen as \(\beta\) and the other one as \(a_1\) or \(b_1\). And for any values of these parameters, the theory has a massless graviton and no other particle around the unique viable vacuum given by the following equation
\begin{equation}
\frac{\lambda}{n-2} - 1 + (\lambda_0 + 1) \left( \frac{\lambda}{n-2} + 1  \right)^{1-n} = 0,
\end{equation}
where \(\lambda = \gamma \Lambda\) is the \textit{dimensionless} effective cosmological constant. Although looking quite cumbersome, remarkably for any \(n\), the vacuum equation gives a unique viable solution consistent with the unitarity requirements. This subtle fact requires a long analysis of the root structure of this equation and we shall not repeat it here as it was done in the Appendix-C of \cite{GulluSismanTekin2}.

As expected, the theory (\ref{L3-2}) reproduces cosmological Einstein's gravity at the first order expansion in curvature; it reproduces Einstein-Gauss-Bonnet theory at the second order expansion and a particular cubic and quartic theory in the next two order expansions. The important point here is that, all these perturbative theories are unitary on their own with the same particle content as Einstein's theory, namely they just have one massless spin\(-2\) excitation about their unique vacuum solution. Moreover, the theory keeps all of its properties at any truncated order in an expansion in powers of the curvature. The form of the vacuum equation and the effective gravitational constant are modified up to and including the \( \mathcal{O}(R^4)  \) expansion but beyond that, namely at \( \mathcal{O}(R^{4 + k})  \) with \(k \geq 1\),  the structure of the vacuum equation or the propagator of the theory are not affected. This is due to the fact that the action is defined with a determinant operation and only up to second order terms in curvature are kept inside the determinant.

The layout of this paper is as follows: In section 2, we define the Wald entropy for a generic theory and carry out the computation of the \(n-\)dimensional Born-Infeld gravity. In section 3, we study a minimal version of BI gravity and consider the \(n \to \infty\) limit of the theory and show that the resulting exponential gravity has the same area-law for the entropy. In section 4, we study the \(3-\)dimensional Born-Infeld gravity which has a massive graviton in its perturbative spectrum.

\section{Wald Entropy in Generic Born-Infeld Gravity}

In this paper, we shall consider the locally (A)dS metrics, but for the sake of generality and possible further work, in this section we consider a general spherically symmetric and static metric and give its curvature properties which will be relevant to black holes as well as (A)dS spacetimes. Therefore we consider the following metric
\begin{equation}
\label{metric}
ds^2 = - a(r)^2 dt^2 + \frac{1}{a(r)^2} dr^2 + \sum_{i,j=2}^{n-2}r^2h_{ij}(x^k)dx^i dx^j,
\end{equation}
where \(h_{ij}(x^k)\) is the induced metric on \((n-2)-\)dimensional sphere. The Christoffel symbols for this metric are
\begin{align}
\label{christoffel}
\Gamma^0{}_{10}=\frac{a'}{a}, \qquad \Gamma^1{}_{00}= a^3a', \qquad \Gamma^1{}_{11}=-\frac{a'}{a}, \qquad \Gamma^1{}_{ij} = -ra^2h_{ij}, \qquad \Gamma^i{}_{1j} = \frac{\delta^i_j}{r},
\end{align}
where \(a\) is the shorthand notation for \( a(r)\) and prime denotes derivative with respect to the radial coordinate \(r\). From the definition of the Riemann tensor
\begin{equation}
R^{\rho}{}_{\sigma \mu \nu} = \partial_{\mu}\Gamma^{\rho}{}_{\nu \sigma} - \partial_{\nu} \Gamma^{\rho}{}_{\mu \sigma} + \Gamma^{\rho}{}_{\mu \lambda} \Gamma^{\lambda}{}_{\nu \sigma} - \Gamma^{\rho}{}_{\nu \lambda} \Gamma^{\lambda}{}_{\mu \sigma},
\end{equation}
and its contractions, the corresponding curvature terms follow
\begin{align}
\label{curvature}
& R_{0101} = (a')^2 + aa'', \qquad R_{0i0j}  = r a^3 a' h_{ij}, \qquad R_{1i1j} = - \frac{r a'}{a} h_{ij},
\nonumber
\\
& R_{ijkl}  = -r^2 (a^2 - 1) (h_{ik}h_{jl} - h_{il}h_{jk}), \qquad R_{00} = a^2 \left( (a')^2 + a a'' + \frac{n-2}{r} a a'  \right), 
\nonumber
\\
& R_{11} = -\frac{1}{a^2} \left( (a')^2 + a a'' + \frac{n-2}{r} a a'  \right), \qquad R_{ij} = - \left[(n-3)(a^2 -1) + 2 r a a')\right] h_{ij},
\nonumber
\\
& R   = - \left( 2(a')^2 + 2 a a'' + \frac{4(n-2)}{r} a a' + \frac{(n-2) (n-3)}{r^2} (a^2 - 1)   \right).
\end{align}

Note that since de Sitter spacetime is maximally symmetric its curvature terms can be written collectively as
\begin{align}
\label{deSitter}
R_{\mu \nu \alpha \beta}  = \frac{2 \Lambda}{(n-1)(n-2)} \left( g_{\mu \alpha}g_{\nu \beta}  - g_{\mu \beta}g_{\nu \alpha}  \right), \qquad
R_{\mu \nu} = \frac{2\Lambda}{n-2} g_{\mu \nu},
\qquad
R = \frac{2 \Lambda n}{n-2}.
\end{align}
In addition to that, the following exact relation will be employed for calculations:
\begin{align}
C_{\mu\rho\sigma\lambda}C_{\nu}{}^{\rho\sigma\lambda}  = \, R_{(\mu}{}^{\rho\sigma\lambda}&R_{\nu) \rho\sigma\lambda}  + \frac{4}{n-2}  R_{(\mu}{}^{\rho \sigma}{}_{\nu)} R_{\rho \sigma} - \frac{2n}{(n-2)^2} R^{\lambda}{}_{(\mu} R_{\nu) \lambda} + \frac{4}{(n-2)^2}R  R_{\mu \nu} 
\nonumber
\\
& + \frac{2}{(n-2)^2} \left( R_{\rho \sigma}^2 - \frac{1}{n-1}R^2    \right)g_{\mu \nu}  ,
\end{align}
whose contraction yields
\begin{equation}
C_{\alpha \beta \rho \sigma}^2   = \, R_{\alpha \beta \rho \sigma}^2 -\dfrac{4}{n-2}R_{\rho \sigma}^2 +\dfrac{2}{(n-1)(n-2)}R^2.
\end{equation}

Now we turn to the computation of Wald entropy for the BI theory (\ref{L3-2}). Recall that \cite{Wald1, Wald2, Wald3, Wald4} Wald entropy arises as a Noether charge of diffeomorphism symmetry of the theory and for a generic action of the form
\begin{equation}
I = \frac{1}{16 \pi G} \int_{\mathcal{M}} d^n x \sqrt{-g} \mathcal{L}\left(g_{ab}, R_{abcd}    \right),
\end{equation}
it reads as \footnote{Note that there is an overall factor between \(L\) and \(\mathcal{L}\), namely \( L = \frac{1}{16 \pi G} \mathcal{L}. \)}
\begin{equation}
\label{waldentropy}
S_W = -2\pi \oint  \left(\frac{\partial L}{\partial R_{abcd}} \varepsilon_{ab} \varepsilon_{cd}\right) dV_{n-2} ,
\end{equation}
where \(r=r_H\) denotes the bifurcate Killing horizon and the integral is to be evaluated on-shell.
The volume element \(dV_{n-2}\) is the volume element on the bifurcation surface \(\Sigma\) and \(\varepsilon_{ab}\) are the binormal vectors to \(\Sigma\). For the metric (\ref{metric}) the binormal vectors can quite easily be found as
\begin{equation}
\varepsilon_{01} = 1, \qquad \varepsilon_{10} = -1,
\end{equation}
by using the timelike Killing vector \(\chi \rightarrow (-1,0,0,\ldots)\) in the formula
\begin{equation}
\varepsilon_{a b} = - \frac{1}{k} \nabla_{a} \chi_{b},
\end{equation} 
where \(k\) is defined as \( k = \sqrt{-\frac{1}{2} (\nabla_{\alpha} \chi_{\beta}) (\nabla^{\alpha} \chi^{\beta}) }   \).

To find the Wald entropy of the de Sitter spacetime which is given by the metric
\begin{equation}
ds^2 = - \left(1 - \frac{2\Lambda r^2}{(n-1)(n-2)}\right) dt^2 + \left(1 - \frac{2\Lambda r^2}{(n-1)(n-2)}\right)^{-1} dr^2 + r^2 d\Omega_{n-2}^2,
\end{equation}
we need to compute the derivative of the Lagrangian which can be done as follows. By defining \( f(\mathsf{A}) =  \sqrt{-\det (g_{\mu \nu}+ \gamma A_{\mu \nu})} \) and using \( \sqrt{\det M_{\mu \nu}} = \exp \left[\frac{1}{2} \text{Tr} \ln M_{\mu \nu}   \right]  \) we obtain
\begin{equation}
\frac{\partial \mathcal{L}}{\partial R_{a b c d}} =  f(\mathsf{A}) \left(\left( \mathsf{g + \gamma A}  \right)^{-1}\right)^{\mu \nu} \frac{\partial A_{\mu \nu}}{\partial R_{abcd}}.
\end{equation}
Here, when calculating the derivative one needs to use 
\begin{align}
\label{derivatives}
\frac{\partial R}{\partial R_{abcd}}  = g^{c[a} g^{b]d},  \qquad  \frac{\partial R_{\mu \nu}^2}{\partial R_{abcd}} = g^{a[c}R^{d]b} - g^{b[c}R^{d]a}, \qquad  \frac{\partial R_{\mu \nu \alpha \beta}^2}{\partial R_{abcd}} = 2  R^{abcd},
\end{align}
and it is a lot easier if \(A_{\mu \nu}\) is written in Riem-Ric-R (RRR) basis as it is done in \cite{GulluSismanTekin1}. By denoting a bar on top of the background elements we can evaluate all the terms in de Sitter background with the help of (\ref{deSitter}). The derivative of the \(\mathsf{A}-\)tensor reads
\begin{align}
	 \left(\frac{\partial A_{\mu \nu}}{\partial R_{abcd}}\right)_{\bar{R}}  \; =  \;  \bigg(\frac{1+\beta}{4} & + \frac{ \lambda}{n-2}a_3   \bigg) \left( \delta^b_{(\mu} \delta^d_{\nu)}\bar{g}^{ac} - \delta^b_{(\mu} \delta^c_{\nu)}\bar{g}^{ad} - \delta^a_{(\mu} \delta^d_{\nu)}\bar{g}^{bc} + \delta^a_{(\mu} \delta^c_{\nu)}\bar{g}^{bd}   \right)
	\nonumber
	\\
	& + \left(-\frac{\beta}{2n} + \frac{2\lambda}{n(n-2) }b_2\right) \left( \bar{g}^{ac}\bar{g}^{bd} - \bar{g}^{ad}\bar{g}^{bc}  \right) \bar{g}_{\mu \nu},  
\end{align}
where \(\lambda = \gamma \Lambda > 0\). The value of the other tensors can be computed as
\begin{equation}
	\bar{g}_{\mu \nu} + \gamma \bar{A}_{\mu \nu}  \; = \;  \bar{g}_{\mu \nu} \left( 1 + \chi  \right),
\end{equation}
where \( \chi = \frac{2\lambda}{n-2} + \frac{4  \lambda^2}{(n-2)^2} (a_3 + b_2)     \) and finally one has
\begin{equation}
	f(\mathsf{A})_{\bar{R}} \;   =  \;   (1 + \chi)^{n/2}.
\end{equation}
Using these with the implementation of the binormal vectors one obtains the Wald entropy for the \(n-\)dimensional de Sitter spacetime as 
\begin{equation}
\label{WaldL3}
S_W =\frac{A_H}{4G} \left( 1 + \frac{4 \lambda}{n-2}(a_3 + b_2) \right) \left( 1 + \frac{2\lambda}{n-2} + \frac{4 \lambda^2}{(n-2)^2} (a_3 + b_2)  \right)^{\frac{n-2}{2}}.
\end{equation}
After inserting the constraints (\ref{constraints}) on \(a_3\) and \(b_2\)  for the unitarity of the theory, Wald entropy takes the form
\begin{equation}
\label{WaldBI}
	S_W = \frac{A_H}{4G} \left(  1 + \frac{\lambda}{n-2}   \right)^{n-1},
\end{equation}
over which two important limits of small and large \(\gamma\) can be taken as
\begin{align}
	\lim_{\gamma \to 0} S_W & = \frac{A_H}{4G} = S_{BH},
	\\
	\lim_{\gamma \to \infty} S_W & = \frac{A_H}{4G} \left(\frac{\Lambda}{n-2}\right)^{n-1}\gamma^{n-1}.
\end{align}
As expected when the Born-Infeld parameter \(\gamma\) is taken to be very small the theory gives the results of Einstein's theory and the Bekenstein-Hawking and Wald results match. In the large \(\gamma-\)limit the crucial result to note is the increase in entropy when \(\gamma\) increases, i.e. when we add more curvature to the spacetime, entropy increases dramatically with the cubic order in four dimensions.

The \textit{effective gravitational constant} of the theory again might be calculated from the derivative of the Lagrangian with respect to Riemann tensor: \footnote{For more details on this see \cite{TahsinSon}. }
\begin{equation}
\frac{1}{\kappa_\text{eff}} = \frac{1}{G} \frac{\bar{R}_{\mu \nu \rho \sigma}}{\bar{R}} \left( \frac{\partial \mathcal{L}}{\partial R_{\mu \nu \rho \sigma}}    \right)_{\bar{R}},
\end{equation}
which in our case gives
\begin{equation}
\label{kappaeff1}
\frac{1}{\kappa_\text{eff}} =  \frac{1}{G}\left( 1 + \frac{\lambda}{n-2}   \right)^{n-1}.
\end{equation}
This result has some important implications since it allows us to rewrite the Wald entropy in a nicer form
\begin{equation}
S_W =\frac{A_H}{4\kappa_{\text{eff}}},
\end{equation}
which is consistent with the results of \cite{Brustein}. Namely, as far as the dynamical entropy is concerned, BI gravity that has the same particle content of Einstein's theory has the Bekenstein-Hawking form. The only change is that instead of \(G\), \(\kappa_{\text{eff}}\) must be used. And since \( \kappa_{\text{eff}} < G \) for dS, as can be seen from (\ref{kappaeff1}), the entropy is increased due to the high curvature terms.

Until one carries out the computations leading to our equations (\ref{WaldL3}) and (\ref{WaldBI}) for Born-Infeld gravity theories that we  defined in our previous papers and the ones in 3 dimensions and infinite number of dimensions, it is absolutely not clear at all that the entropy of the de Sitter Space in these theories obey the area law.  This is actually quite a surprise since almost any other theory with some powers of curvature will not have this property even for the de Sitter case. Here the main idea is to see a possible connection between the particle content, vacuum and entropy properties of a theory at hand.  The Born-Infeld type theories that we have constructed before are in some sense very natural generalizations of Einstein's gravity as  far as their  particle content and vacua are concerned. In this work we have shown another shared property of these models: that is  both Einstein's theory and the specific BI gravity theories that we discussed obey the area law for entropy. This statement is important as it shows the equivalence of the dynamical and geometric entropies in these models.

\section{Two Special BI gravities}

Observe that the Wald entropy (\ref{WaldBI}) does not depend on the parameters \(\beta\), \(a_3\) and \(b_2\). Namely, these parameters are not constrained further by the positivity of the Wald entropy. Obeying the constraints they must satisfy, one can write down a \textit{minimal} theory without any parameters which reads as \cite{GulluSismanTekin1}
\begin{equation}
\label{BI}
\mathcal{L}= \frac{2}{\gamma}  \left[ \left(1+\frac{\gamma}{n}R+\frac{\gamma^2(n-1)^2}{4n(n-2)(n-3)}\chi_{GB}-\frac{\gamma^2(n-2)}{4n^2}R^2\right)^{n/2}-\lambda_{0}-1  \right],
\end{equation}
where $\chi_{GB}=R_{\mu \nu \alpha \beta}^2-4R_{\mu \nu}^2+R^2$ is the Gauss-Bonnet term and the Wald entropy for the dS spacetime is still given by (\ref{WaldBI}). This theory has further studied in \cite{Esin} where it was shown that the Schwarzschild-Tangherlini black hole is only an approximate solution.

General relativity in the large number of dimensions has some interesting properties (see \cite{Emparan}). Here we can also consider the \(n \to \infty\) limit of the BI gravity given by (\ref{BI}) to get \cite{GulluSismanTekin1}
\begin{equation}
\mathcal{L}  = \frac{2}{\gamma} \left(  \exp{\left[\frac{\gamma}{2}R + \frac{\gamma^2}{8} \Big( R_{\mu \nu \alpha \beta}^2- 4 R_{\mu \nu}^2 \Big)\right]} - \lambda_0 - 1 \right),
\end{equation}
where the Wald entropy reads as
\begin{equation}
	S_W = \frac{A_H}{4G} \left( 1 - \frac{2 \lambda (2n-3)}{(n-1)(n-2)} \right) \exp \left[ \frac{\lambda n }{n-2} \left( 1 - \frac{\lambda (2n-3)}{(n-1)(n-2)}    \right)   \right].
\end{equation}

Here, in addition to the small\(-\gamma\) limit 
\begin{equation}
	\lim_{\gamma \to 0} S_W  =  \frac{A_H}{4G} = S_{BH},
\end{equation}
which gives the expected result and reduces to the Bekenstein-Hawking entropy, one can also check the large\(-n\) limit
\begin{equation}
	\lim_{n \to \infty} S_W  = \frac{A_H}{4G} e^{\lambda},
\end{equation}
that satisfies the relation \(S_W = \frac{A_H}{4 \kappa_\text{eff}}  \) where \(  \kappa_{\text{eff}}\) can be found as (or can be obtained from (\ref{kappaeff1}) by taking the \(n \to \infty \) limit ) \( \frac{1}{\kappa_{\text{eff}}} = \frac{e^{\lambda}}{G}   \).

\section{Born-Infeld extension of New Massive Gravity}

As we noted in the Introduction, the Born-Infeld gravity in \(2+1\) dimensions has a particularly simple form which was obtained as an infinite order expansion of the quadratic new massive gravity \cite{Bergshoeff, Bergshoeff2, GulluSismanTekin3}. BI theory has the nice property that it coincides at any order in curvature expansion with the theory obtained from the requirements of AdS/CFT \cite{Sinha, Paulos} and moreover it appears as a counter term in AdS\(_4\) \cite{Jatkar}. Both BI and new massive gravity describe a massive spin\(-2\) particle with \(2\) degrees of freedom in \(2+1\) dimensions and have very rich solution structures, including the BTZ and various other black holes. For further work on this theory see \cite{Alishahiha, Ghodsi}.

For the theory given in (\ref{3d_action}), we can define the Wald entropy of the horizon of the BTZ black hole in terms of the derivative of the Lagrangian with respect to the Ricci tensor as
\begin{equation}
	S_W = -2\pi \oint  \left(\frac{\partial L}{\partial R_{\rho \sigma}} g^{\mu \nu} \varepsilon_{\mu \rho} \varepsilon_{\nu \sigma}\right) d^3x.
\end{equation}
To carry out the computation, let us define
\begin{equation}
	f(R_{\mu \nu}) = \sqrt{-\det \left( \mathsf{g} + \frac{\sigma}{m^2} \mathsf{G}  \right)},
\end{equation} 
which yields
\begin{equation}
	\frac{\partial f}{\partial R_{\rho \sigma}} = \frac{\sigma}{4m^2} f \left(\left( \mathsf{g} + \frac{\sigma}{m^2} \mathsf{G}  \right)^{-1}\right)^{\mu \nu} \left( \delta^{\rho}_{\mu} \delta^{\sigma}_{\nu} + \delta^{\rho}_{\nu} \delta^{\sigma}_{\mu} - g_{\mu \nu} g^{\rho \sigma} \right),
\end{equation}
where the derivative of the Ricci tensor is taken by respecting its symmetry, \( \frac{\partial R_{\mu \nu}}{\partial R_{\rho \sigma}} = \frac{1}{2} \left(   \delta^{\rho}_{\mu} \delta^{\sigma}_{\nu} + \delta^{\rho}_{\nu} \delta^{\sigma}_{\mu}   \right)    \).

Locally the BTZ solution is AdS with the curvature values
\begin{equation}
	R_{\mu \rho \nu \sigma} = \Lambda \left( g_{\mu \nu} g_{\rho \sigma} - g_{\mu \sigma} g_{\rho \nu}   \right), \qquad R_{\rho \sigma} = 2 \Lambda g_{\rho \sigma}, \qquad R = 6 \Lambda,
\end{equation}
and from here one can calculate the following relevant quantities
\begin{align}
	\mathsf{\bar{g}} + \frac{1}{m^2} \mathsf{\bar{G}} & \; = \; \left( 1 - \frac{\sigma \Lambda}{m^2}   \right) \mathsf{\bar{g}},
	\\
	f(R_{\mu \nu})_{\bar{R}}& \; = \; \sqrt{- \bar{g}} (1 - \frac{\sigma \Lambda}{m^2})^{3/2},
\end{align}
which leads to \footnote{Note that \( \sqrt{-g} \) appearing in \(f(R_{\mu \nu})\) is promoted to integral density as was done in (\ref{L3-2}). }
\begin{equation}
\left(\frac{\partial L}{\partial R_{\rho \sigma}}\right)_{\bar{R}}  =  \frac{\sigma}{16\pi G_3} \sqrt{1 - \frac{\sigma \Lambda}{m^2}} \; \bar{g}^{\rho \sigma}.
\end{equation} 
Since the binormal vectors are the same as before, one arrives at
\begin{equation}
	S_W = \frac{A_H}{4} \frac{\sigma}{G_3} \sqrt{1 - \frac{\sigma \Lambda}{m^2}},
\end{equation}
which is positive for \(\sigma = +1\). This result matches \cite{Nam}. To put this result into the Bekenstein-Hawking form, let us compute the effective gravitational constant which reads
\begin{equation}
	\frac{1}{\kappa_\text{eff}} = \frac{1}{G_3}\frac{\bar{R}_{\mu \nu }}{\bar{R}} \left( \frac{\partial \mathcal{L}}{\partial R_{\mu \nu}}    \right)_{\bar{R}} =  \frac{\sigma}{G_3} \sqrt{1 - \frac{\sigma \Lambda}{m^2}},
\end{equation}
leading to the relation \(S_W = \frac{A_H}{4 \kappa_\text{eff}}\). Since \(\Lambda <0\), entropy is again increased due to the higher curvature terms as in the generic Born-Infeld gravity.

As it is noted in \cite{BINMG1}, there is another theory that can be considered as an extension of the new massive gravity whose action reads
\begin{equation}
	I = -\frac{m^2}{4\pi G_3} \int d^3x \left( \sqrt{-\det \left[ g_{\mu \nu} + \frac{\sigma}{m^2 } \left( R_{\mu \nu} - \frac{1}{6}g_{\mu \nu} R  \right)    \right]}  - \left( 1 - \frac{\lambda_0}{2}    \right) \sqrt{-  g} \right).
\end{equation} 
By carrying out the analogous computations for this action one arrives at the Wald entropy of the BTZ solution as
\begin{equation}
	S_W  = - \frac{A_H}{4} \frac{\sigma}{G_3}\sqrt{1 + \frac{\sigma \Lambda}{m^2}},
\end{equation}
where this time entropy is positive for \(\sigma = -1\) and the relation \(S_W = \frac{A_H}{4 \kappa_\text{eff}}\) is again satisfied with effective gravitational constant calculated as \( \frac{1}{\kappa_\text{eff}} = -\frac{\sigma}{G_3} \sqrt{1 + \frac{\sigma \Lambda}{m^2}} \). 

\section{Conclusions}

We have studied the Wald entropy for Born-Infeld type gravity theories which were constructed to carry the vacuum and excitation properties of Einstein's gravity. Namely, these BI theories have a unique maximally symmetric solution (despite the many powers of curvature) and a massless spin\(-2\) graviton about the vacuum for \(n \geq 3 + 1\) dimensions (and a massive graviton for \(n=2+1\) dimensions). Wald entropy (dynamical entropy coming from diffeomorphism invariance) usually differs from the Bekenstein-Hawking entropy for generic gravity theories. But here we have shown that these two notions of entropy coincide for BI gravity theories in generic dimensions with the slight modification that instead of the Newton's constant of Einstein's theory, the effective gravitational coupling appears. Basically the following expression is valid:
\begin{equation}
	S_{W} = S_{BH} = \frac{A_H}{4 \kappa_{\text{eff}}},
\end{equation}
where \(\kappa_{\text{eff}}\) encodes all the information about the underlying theory as far as its de Sitter entropy is concerned. Hence, besides having similar properties at the vacuum and linearized levels, these theories share the same entropy properties for their de Sitter solutions. This is a rather interesting result and one might conjecture that it might be a property of gravity theories that have only massless gravity in their spectrum. (The \(3-\)dimensional case is somewhat different as there is no massless bulk graviton and one necessarily considers massive gravity). To be able to understand this potentially deep connection between various notions of entropy and the particle content of the underlying gravity theory, further solutions of BI gravity theories must be constructed. This work is underway.

\section*{Acknowledgment}
S.K. is supported by the TUBITAK 2211-A Scholarship.

\appendix

\section{A Brief Look at Black Hole Entropy in Born-Infeld Gravity}

In the bulk of the paper we studied the entropy of the de Sitter space for generic gravity and did not discuss black hole entropy (except for the \(n=2+1\) dimensions). The main reason for this is that the BI theories studied here are quite recent and no exact black hole type solutions have been found for these theories. The only relevant work up to now is \cite{Esin} where perturbative solutions were constructed. But for the sake of completeness and as an initial estimate to what happens to the entropy of Schwarzschild-Tangherlini black holes in the BI gravity, let us carry out the following computation which should be considered as a preliminary work. To be specific, let us take the Lagrangian (\ref{BI}) which reads
\begin{equation}
\label{Lappendix}
\mathcal{L}= \frac{2}{\gamma}\left( f(R_{\mu \nu \alpha \beta})^{n/2} - \lambda_{0}-1 \right),
\end{equation}
where
\begin{align}
f(R_{\mu \nu \alpha \beta}) = 1+  p R + u R_{\mu \nu \alpha \beta}^2 - 4 u R_{\mu \nu}^2 + (u-v) R^2,
\end{align}
with the coefficients
\begin{equation}
	p=\frac{\gamma}{n}, \quad u=\frac{\gamma^2 (n-1)^2}{4n(n-2)(n-3)}, \quad	v=\frac{\gamma^2 (n-2)}{4n^2}.
\end{equation}
With the help of (\ref{derivatives}), the derivative of the Lagrangian with respect to the Riemann tensor can be found as
\begin{align}
\frac{\partial \mathcal{L}}{\partial R_{abcd}}  =  \frac{n}{\gamma} f(R_{\mu \nu \alpha \beta})^{\frac{n-2}{2}} \Bigg[ \Big[p + 2(u-v)R \Big] g^{c[a} g^{b]d} + 2 u R^{abcd} - 4 u \left( g^{a[c}R^{d]b} - g^{b[c}R^{d]a}    \right) \Bigg]. 
\end{align}
Initially, we want to keep everything general by studying the \(n-\)dimensional Schwarzschild-Tangherlini black hole metric
\begin{equation}
ds^2 = - \left(1 - \frac{c}{r^{n-3}}\right) dt^2 + \left(1 - \frac{c}{r^{n-3}}\right)^{-1} dr^2 + r^2 d\Omega_{n-2}^2,
\end{equation} 
where \(r = r_H = c^{\frac{1}{n-3}}\) is the radius of the black hole horizon. After making a general derivation we will study the \(n=4\) case.

With some effort, one can find the relevant curvature tensor components and make the following table:

\vspace{5mm}

\begin{table}[h]
	\centering
	\begin{tabular}{ p{3cm}||p{6cm}|p{4cm}  }
		\hline \hline
		\multicolumn{3}{c}{Black Hole, \( a = \sqrt{1- \frac{c}{r^{n-3}}}  \)}  \\ 
		\hline \hline
		Curvature Terms & \(n-\)dimensions & \(n=4, c=2M\) \\
		\hline \hline
		\(\boldsymbol{R_{0101}}\)  & \( - \frac{(n-2)(n-3)}{2r^2} \frac{c}{r^{n-3}}   \)    & \( - \frac{2M}{r^3}  \)   \\
		\(\boldsymbol{R_{0i0j}}\)  &   \( \frac{n-3}{2} \frac{c}{r^{n-1}} \left( 1 - \frac{c}{r^{n-3}}   \right) g_{ij} \)  & \( \frac{M}{r^3} \left( 1 - \frac{2M}{r}  \right) g_{ij}  \) \\
		\(\boldsymbol{R_{1i1j}} \) & \( -\frac{n-3}{2} \frac{c}{r^{n-1}} \left( 1 - \frac{c}{r^{n-3}}   \right)^{-1} g_{ij} \)&  \( -\frac{M}{r^3} \left( 1 - \frac{2M}{r}  \right)^{-1} g_{ij}  \)  \\
		\(\boldsymbol{R_{ijkl}}\) & \( \frac{c}{r^{n-1}}\left( g_{ik}g_{jl} - g_{il}g_{jk}   \right)  \) & \( \frac{2M}{r^3}\left( g_{ik}g_{jl} - g_{il}g_{jk}   \right)  \) \\
		\hline
		\( \boldsymbol{R^2_{\mu \nu \alpha \beta}} \) & \(\frac{(n-1)(n-2)^2(n-3)}{r^4} \left( \frac{c}{r^{n-3}}  \right)^2\) & \( \frac{48 M^2 }{r^6} \) \\
	\end{tabular}
	\caption{List of non-zero curvature terms for the black hole solutions in both generic \(n-\)dimensions and also in \(n=4\). Note that all the Ricci tensor components and hence the Ricci scalar vanishes.}
	\label{table:1}
\end{table}

\vspace{5mm}

Having all these information one can obtain the Wald entropy in \(n-\)dimensions for (\ref{Lappendix}) as
\begin{equation}
\label{Waldn}
S_W  =    \frac{ A_H}{4G} \left( 1 + \frac{ \gamma (n-1)^2 }{2 r_H^2} \right) \left( 1 + \frac{\gamma^2 (n-2)(n-1)^3}{4 n r_H^4}  \right)^{\frac{n-2}{2}}.
\end{equation}
In four dimensions, where we have \(  c = 2M, \;  a = \sqrt{1 - \frac{2M}{r}}   \), the result reduces to
\begin{equation}
\label{Wald4}
S_W    = \frac{A_H}{4G} \left( 1 + \frac{27 \gamma^2 }{ 128 M^4}  \right) \left( 1 + \frac{9\gamma }{8 M^2}  \right).
\end{equation}
What we learn from (\ref{Waldn}) and (\ref{Wald4}) is that in the BI gravity, the entropy of the spherically symmetric black holes increases compared to the Einstenian result. We also see that as \( \gamma \to 0 \), one recovers the Bekenstein-Hawking entropy. It is quite possible that one can write \(S_W  \) as the geometric entropy with a modified gravitational constant. But for this to work out, we need the exact black hole solutions in the BI gravity which we currently lack. This is an outstanding problem.

\end{document}